\documentstyle[prl,preprint,aps]{revtex}

\begin{document}
\bibliographystyle{prsty}
\title{Frequency locking of tunable diode lasers to a 
stabilized ring-cavity resonator}
\author{Ayan Banerjee, Dipankar Das, Umakant D. Rapol, and 
Vasant Natarajan \cite{email}}
\address{Department of Physics, Indian Institute of Science, 
Bangalore 560 012, INDIA}

\maketitle
\begin{abstract}
We demonstrate a technique for locking the frequency of a 
tunable diode laser to a ring-cavity resonator. The resonator is 
stabilized to a diode laser which is in turn locked to an 
atomic transition, thus giving it absolute frequency 
calibration. The ring-cavity design has the principal 
advantage that there is no feedback destabilization of the 
laser. The cavity has a free-spectral range of 1.3 GHz and 
$Q$ of about 35, which gives robust locking of the laser. 
The locked laser is able to track large scans of the cavity.
\end{abstract}

The easy availability of diode lasers over most of the 
near-infrared spectrum has led to their widespread use in atomic 
and optical physics experiments \cite{WIH91}. The principal 
advantages of these lasers lie in their low cost, narrow 
spectral width, tunability over several nm, efficient power 
consumption, reliability, and ease of operation. In 
addition, by placing a single-mode diode in an external 
cavity and using optical feedback from an angle-tuned 
grating, it can be made to operate at single frequency 
(single longitudinal mode) with linewidth of order 1 MHz 
\cite{MSW92,RWE95}. Such frequency-stabilized diode lasers 
find applications in several experiments as low-cost 
replacements for ring-cavity dye or Ti-sapphire lasers. In 
particular, experiments on laser cooling have become 
accessible to a large community of researchers because 
several alkali atoms and alkali-like ions have cooling 
transitions which are accessible with diode lasers 
\cite{nobel97}. However, to use lasers in such applications 
effectively it is necessary to set their absolute frequency 
with $\sim$MHz precision. In many cases, this is achieved by 
locking the laser to a known atomic or molecular transition 
using saturated-absorption spectroscopy in a vapor cell. But 
this may not always be possible, especially when working 
with short-lived radioactive species or trapped ions. 

In this Letter, we solve this problem by locking a tunable 
diode laser to a stabilized ring-cavity resonator. Tunable 
lasers are traditionally locked to linear Fabry-Perot 
reference cavities. In fact, several commercial dye and 
Ti-sapphire lasers use a temperature-controlled linear cavity 
for locking. The main problem with such a design for the 
diode laser is that it can cause unwanted feedback into the 
laser and destabilize it. By contrast, our ring-cavity 
design has a traveling wave inside the cavity and there is 
no possibility of feedback into the laser. The ring cavity 
has several other advantages: i) The cavity provides at 
least three output beams that can be used to measure the 
signal from the cavity. ii) The angle of incidence on the 
cavity mirrors is not normal, therefore the mode structure 
inside the cavity is elliptical; diode laser outputs are 
also elliptical, and this makes it easier to mode-match the 
output of the diode laser into the cavity. iii) The design 
is compact and the cavity is easily temperature controlled 
to increase its passive stability.

The cavity is actively stabilized to a diode laser (called 
the ``master laser'') that is itself locked to an atomic 
transition. Using a diode laser locked to an atomic 
transition as the absolute frequency reference has several 
advantages over alternatives such as using a stabilized HeNe 
laser. The atomic transition typically has several hyperfine 
peaks and the laser can be locked to any of them. For 
example, we use the $D_2$ line in Rb for locking the master 
laser. Rb has two stable isotopes, and two ground-state 
hyperfine levels and four excited-state hyperfine levels in 
each isotope. Thus, there are potentially 24 lock points 
available in the saturated-absorption spectrum \cite{foot1}, 
with frequencies varying over a range of several 100 MHz. 
Now, when the arbitrary frequency of a tunable laser (called 
the ``slave laser'') has to be set, its frequency will 
generally be offset from the cavity resonance, and the 
offset can be as large as the cavity free-spectral range. 
This offset is accounted for by shifting the laser frequency 
using an acousto-optic modulator (AOM) and then locking to 
the cavity. If the cavity is stabilized to a single 
frequency from a HeNe laser, the free-spectral range of the 
cavity has to be quite small so that the frequency offset 
lies within the range of the AOM, which is typically limited 
to 100--200 MHz. This implies the use of long cavities with 
their concomitant difficulty in maintaining temperature 
stability. By contrast, the wide range of lock frequencies 
available with the atomic transition means that the cavity 
free-spectral range can be of order GHz, and it is still 
possible to find one lock point for which the slave laser 
frequency lies within the range of the AOM. One other 
advantage of using a diode laser locked to an atomic 
transition is that it makes scanning of the slave laser 
quite simple. The master laser can be scanned over a large 
frequency range around the atomic transition by changing its 
grating angle, and the slave laser tracks this scan. This 
kind of scanning is quite important in spectroscopy 
experiments.

The schematic of our system is shown in Fig.\ 1. The master 
laser is tuned to the $D_2$ line of atomic Rb at 780 nm 
($5S_{1/2} \leftrightarrow 5P_{3/2}$ transition). It is 
locked to one of the hyperfine peaks or crossover resonances 
in the saturated-absorption spectrum. The absolute frequency 
of the Rb $D_2$ line has been measured previously with kHz 
accuracy \cite{YSJ96}, therefore the frequency of this laser 
is known very precisely. The error signal for the locking is 
obtained by modulating the injection current into the diode. 
Next, the cavity is stabilized by locking it to the master 
laser frequency. The error signal for this is obtained by 
modulating the cavity length using a piezo-mounted mirror. 
Finally, the slave laser is locked to a cavity resonance by 
dithering its injection current. All the modulation 
frequencies are in the range of 10--50 kHz and the lock-in 
amplifiers are home built around the Analog Devices AD630 
mixer chip.

The two diode laser systems are built around commercial 
single-mode laser diodes. The output is first collimated 
using a 4.5 mm, 0.55 NA aspheric lens. The laser is then 
frequency stabilized in a standard external-cavity design 
(Littrow configuration) \cite{MSW92} using optical feedback 
from an 1800 lines/mm diffraction grating mounted on a 
piezoelectric transducer. For the master laser, a part of 
the output beam is tapped for Doppler-free 
saturated-absorption spectroscopy in a Rb vapor cell. The laser is 
locked to one of the hyperfine transitions in the $D_2$ line 
by feeding the error signal to the piezoelectric transducer 
that controls the grating angle. The slave laser is tuned 
near the $D_1$ line of Rb at 795 nm.

The ring cavity consists of two plane mirrors and two 
concave mirrors in a bow-tie arrangement, as shown in Fig.\ 
1. One of the plane mirrors is partially reflecting (97\%) 
and is used to couple light into the cavity. The second 
plane mirror is mounted on a piezoelectric transducer and is 
used to adjust the cavity length electronically. The two 
concave mirrors have radius of curvature of 25 mm. The 
concave mirrors are placed 26.5 mm apart, while the optical 
path length between them through the plane mirrors is 200 
mm. The angle of incidence on all mirrors is 15$^{\rm o}$. 
The mirrors are mounted on a 10-mm thick copper plate that is 
temperature controlled to $\pm 0.01^{\rm o}$C using a 
thermoelectric cooler.

We have analyzed the cavity using standard $ABCD$ matrices 
for Gaussian beam propagation \cite{KOL66}. Because of the 
non-zero angle of incidence on the curved mirrors, it is 
necessary to analyze the sagittal and tangential planes 
separately. The cavity modes are therefore elliptical. The 
cavity has two beam waists, a small one of diameter 13.8 
$\mu$m $\times 13.4$ $\mu$m between the concave mirrors and 
a large one of diameter 224 $\mu$m $\times 119$ $\mu$m 
between the plane mirrors. The larger waist is used for 
efficient mode matching of the laser beams into the cavity. 
The output of the diode lasers is directly fed into the 
cavity through a lens of focal length 50 cm. About 15\% of 
the light gets coupled into the cavity. Using an anamorphic 
prism pair after the diode laser, it is possible to adjust 
the ellipticity of the input beam and improve the coupling 
efficiency to 50\%, but we do not do this on a routine 
basis.

The cavity has been designed to have a $Q$ of about 30. The 
measured cavity modes are shown in Fig.\ 2 which is a plot 
of the power inside the cavity as a function of the cavity 
length. The measured $Q$ is 35, which is close to the 
designed value. The free spectral range is 1.3 GHz 
corresponding to a cavity length of 226.5 mm. This implies 
that the cavity modes have a linewidth of about 37 MHz. The 
low $Q$ of the cavity ensures that the locking of the slave 
laser is robust and is quite insensitive to sudden 
perturbations. The error signal obtained by modulating the 
cavity length is fed to the piezo-mounted mirror to lock the 
cavity to the master laser. The piezoelectric transducer has 
a full deflection of 6.1 $\mu$m for a voltage of 150 V. This 
gives a wide dynamic range for the cavity to track scans of 
the master laser.

There are two ways to produce the error signal for locking 
the slave laser to the stabilized cavity. The first 
technique, which is easier to implement, is to modulate the 
injection current into the slave laser. The second technique 
is to modulate the frequency of the AOM which is used to 
shift the frequency of the laser and match it to a cavity 
resonance. The second technique has the advantage that the 
laser frequency does not vary and there is no increase in 
its linewidth. However, the scheme is complicated because it 
requires double passing through the AOM in order to maintain 
directional stability. Therefore, we have used the first 
technique in these experiments. 

In order to test the ability of the slave laser to track 
scans of the master laser, we have used the slave laser to 
measure transitions on the Rb $D_1$ line ($5S_{1/2} 
\leftrightarrow 5P_{1/2}$ transition) while it was locked to 
the cavity. The slave laser is first tuned near the $D_1$ 
line by measuring its frequency using a home-built wavemeter 
\cite{BRW01}. The wavemeter also uses the master laser 
locked to the atomic transition for absolute frequency 
calibration. This gives us the slave-laser frequency with an 
accuracy of about 20 MHz. The frequency shift in the AOM is 
then adjusted so that the slave laser is on a cavity 
resonance and the laser is then locked to the cavity. The 
master laser is now scanned over a frequency range of 
several 100 MHz. The saturated-absorption spectrum of the 
$5S_{1/2},F=2 \rightarrow 5P_{1/2},F'$ transitions in 
$^{85}$Rb is shown in Fig.\ 3. We see that the slave laser 
is able to track scans of order GHz or more without losing lock.

In conclusion, we have shown that the frequency of a tunable 
diode laser can be locked using a stabilized ring-cavity 
resonator. The ring cavity has several advantages over a 
linear etalon, the main one being that there is no feedback 
destabilization of the laser. The cavity is stabilized using 
a diode laser which is in turn locked to an atomic 
transition. The atomic transition has several peaks in the 
spectrum, and thus provides several calibrated frequency 
markers for locking. The cavity we have used in the current 
experiments has a free spectral range of 1.3 GHz and a $Q$ 
of 35. The low $Q$ allows robust locking of the slave laser. 
The piezo-mounted mirror in the cavity has a wide dynamic 
range that allows large scans of the master laser to be 
tracked by the slave laser. We have verified that the slave 
laser stays locked while the master laser is scanned over a 
frequency range of 1 GHz or more. The cavity modes have a 
linewidth of about 37 MHz and this determines the tightness 
of the lock. In future, we plan to use a longer cavity with 
free spectral range of about 500 MHz and mode linewidth of 
15 MHz. We expect to get much tighter locking with the new 
design.

This work was supported by research grants from the Board of 
Research in Nuclear Sciences (DAE), and the Department of 
Science and Technology, Government of India.

\begin{figure}
\caption{
Schematic of the system. The master laser is a 
frequency-stabilized diode laser locked to an atomic transition in Rb. 
The ring cavity has a bow-tie design with two plane and two 
curved mirrors. It is stabilized to the master laser by 
adjusting its length using the piezo-mounted mirror. The 
slave laser is a frequency-stabilized diode laser that is 
locked to the cavity. The acousto-optic modulator (AOM) is 
used to account for any difference between the cavity resonance 
and the slave-laser frequency. Figure key: LIA is Lock-in 
amplifier, PBS is polarizing beamsplitter, M is mirror, BS is
beamsplitter, PD is photodiode.
}
\end{figure}

\begin{figure}
\caption{
Cavity modes. The figure shows the power inside the cavity 
as the cavity length is scanned using the piezo-mounted 
mirror. The mean cavity length $L_0$ is about 226.5 mm. The 
dotted line is a Lorentzian fit to the two peaks which 
yields a $Q$ of 35.
}
\end{figure}

\begin{figure}
\caption{
Scan of the slave laser. The plot is a saturated-absorption 
spectrum of the $5S_{1/2},F=2 \rightarrow 5P_{1/2},F'$ 
transitions in $^{85}$Rb obtained by scanning the master 
laser while the slave laser was locked to the cavity. The 
outer peaks are the hyperfine transitions corresponding to 
$F'=2$ and $F'=3$, while the middle peak is the crossover 
resonance. Probe detuning is measured from the $F=2 \rightarrow
F'=2$ transition. The total scan width is 1.2 GHz.
}
\end{figure}


\begin{references}
\bibitem{foot1}
Each ground level can couple to three excited levels 
producing three peaks in the spectrum. Each pair of peaks 
has a corresponding crossover resonance, thus giving 6 peaks 
per ground level for each isotope.
\end{references}
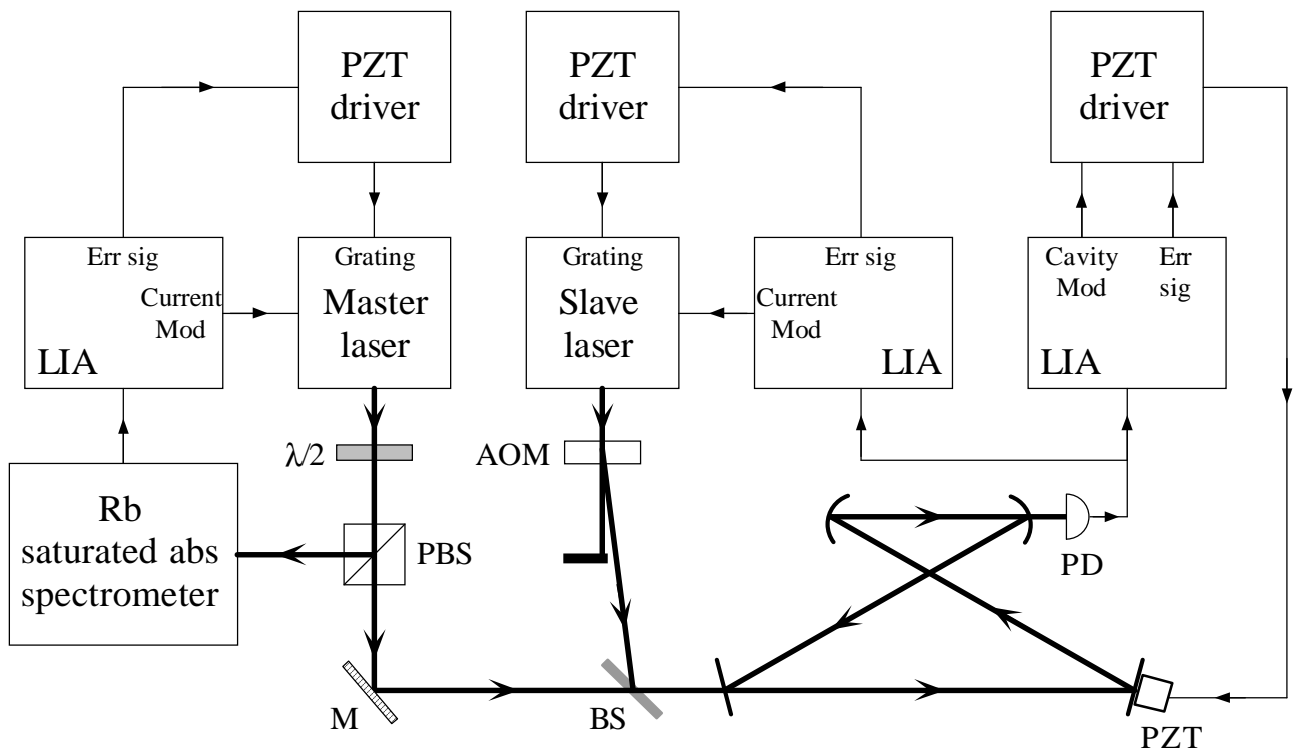
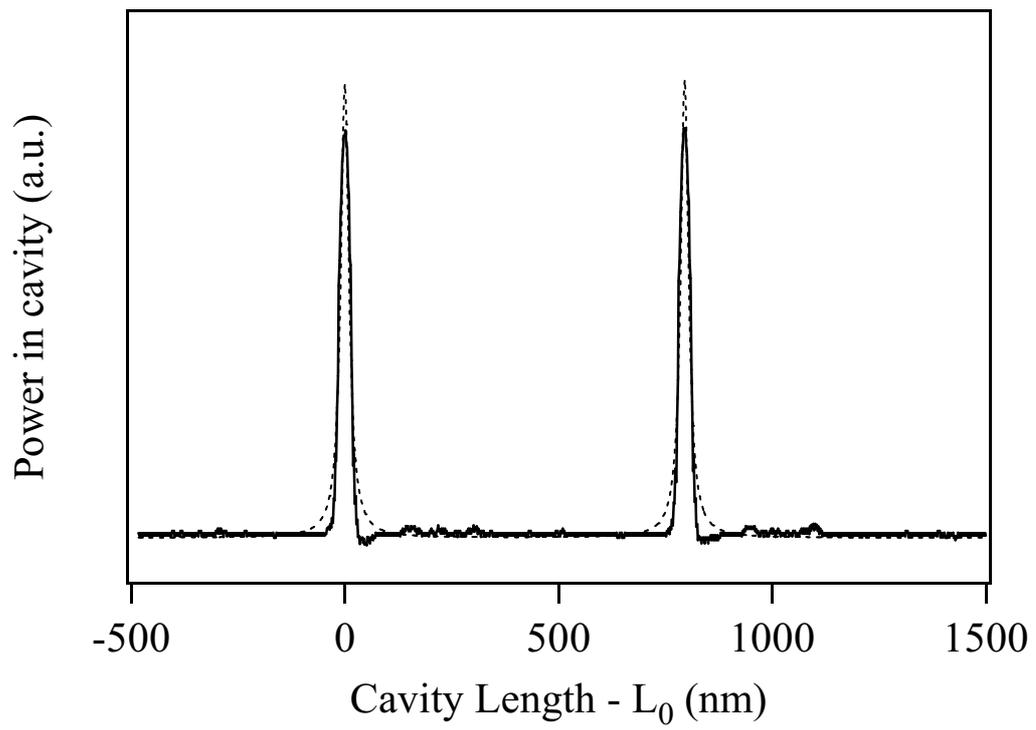
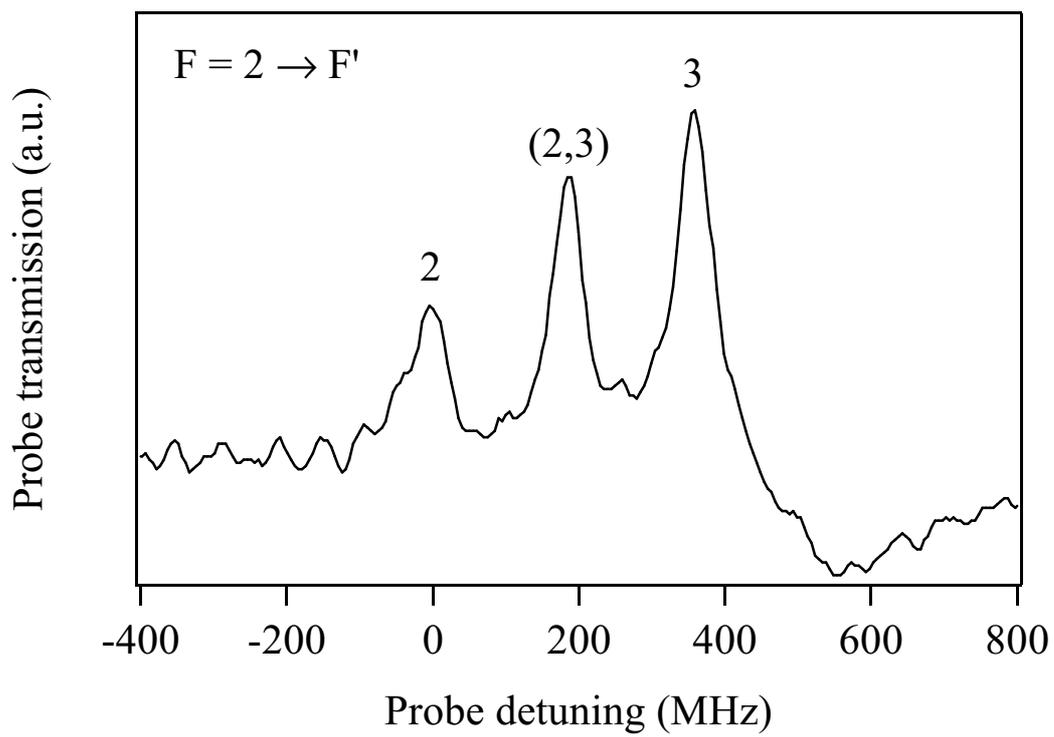
\end{document}